\documentclass[aps,prd,tighten,11pt,superscriptaddress,a4,nofootinbib]{revtex4}

\usepackage[utf8]{inputenc}
\usepackage{graphicx}
\usepackage{bm}
\usepackage{amsmath,amssymb}
\usepackage{latexsym}
\usepackage{color}
\usepackage{tabularx}

\newcommand{\bee}{\begin{equation}}
\newcommand{\eee}{\end{equation}}

\begin{document}

\title{Quasinormal modes of black holes. II. Pad\'e summation of the higher-order WKB terms.}

\author{Jerzy Matyjasek}\email{jurek@kft.umcs.lublin.pl, jirinek@gmail.com}
\affiliation{Institute of Physics,
Maria Curie-Sk\l odowska University\\
pl. Marii Curie-Sk\l odowskiej 1,
20-031 Lublin, Poland} 

\author{Ma{\l}gorzata Telecka}

\affiliation{Faculty of Earth Sciences and Spatial Management,\\ Maria Curie-Sk\l odowska University, \\
Al. Kraśnicka 2cd, 20-031 Lublin, Poland}

\begin{abstract}
In previous work~\cite{ja} we proposed an improvement of the WKB-based semianalytic 
technique of Iyer and Will for calculation of the quasiormal modes of black holes by
constructing the Pad\'e approximants of the formal series for $\omega^{2}.$ It has been
demonstrated that (within the domain of applicability) the diagonal Pad\'e transforms 
$\mathcal{P}_{6}^{6}$ and $\mathcal{P}_{7}^{6}$  are always in a very good agreement 
with the numerical results.  In this paper we present a further extension of the method. 
We show that it is possible to reproduce many known numerical results with a great
accuracy (or even exactly) if the Pad\'e transforms are constructed from the perturbative
series of a really high order. In our calculations the order depends on the problem but
it never exceeds 700. For example, the frequencies of the gravitational mode $l=2,$ $n=0$
calculated with the aid of the Pad\'e approximants and within the framework of the 
continued fractions method agree to 24 decimal places. The use of such a large number of 
terms is necessary as the stabilization of the quasinormal frequencies can be slow. 
Our results reveal some unexpected features of the WKB-based approximations and may shed
some fresh light on the problem of overtones. 
\end{abstract}
\maketitle

\section{Introduction}

The quasinormal modes of black holes~\cite{Vish} have been the subject of intense studies over
the last 40 years and the reason for this continuous interest stems from the fact that they were
expected to be detectable. Indeed, a perturbed black hole undergoes the ringdown phase in which
the oscillations are characterized by a set of complex frequencies, $\omega.$ The real part of
each $\omega$ gives the oscillation frequency whereas the imaginary part determines the
characteristic damping rate and consequently the quasinormal modes are crucial in detecting and
subsequently studying the properties of the gravitational waves generated in the violent
collisions of black holes or gravitational collapse. Because of the constant and still growing
interest in the quasinormal modes many numerical and analytical approaches have been proposed
for their calculation. The most popular are the highly accurate numerical methods, such as the
method of continued fractions~\cite{Leaver1,nollert2,Rostworowski} constructed by applying the Frobenius
series solution, asymptotic iteration~\cite{Nay}, the method of the Hill 
determinant~\cite{Pancha}, the method of Nollert and Schmidt~\cite{nollert1} and the pseudospectral 
method~\cite{Jansen}. Recently, in a very interesting development~\cite{cardoso1,cardoso2},
the quasinormal modes of the
parametrized black holes have been studied.
On the opposite side we have a bunch of the approximate analytic and semianalytic methods 
mainly based on the WKB expansion and its 
variants~\cite{wkb0,wkb1,wkb2,wkb3,wkb4,phase1,phase2}, the perturbative approach of 
Ref.~\cite{OlegZ} and the approach advocated by Gal'tsov and Matiukhin~\cite{Galtsov}.
Especially interesting is the method developed by Zaslavskii~\cite{OlegZ}, who following ideas of 
Refs.~\cite{blome,val1,val2} reduced the problem to the quantum anharmonic oscillator and used
powerful Rayleigh-Schr\"odinger perturbation theory. 

We assume that the second order differential equation describing perturbations (with a time 
dependence of the form $e^{-\omega t}$) can be reduced to the Schr\"odinger-type equation
 \begin{equation}
 \left[\frac{d^{2}}{dx^{2}}  + Q(x)\right] \psi(x) =0,
 \label{schred}
\end{equation}
where $Q(x) = \omega^{2} -V(x),$  $V(x)$ is a potential and $x$ is the tortoise coordinate. 
The potential is constant as
$|x|\to \infty$ ($V(-\infty)$ and $V(\infty)$  may be different) and has maximum at  $x_{0}.$  
The radial part of the free oscillations are described by the functions $\psi(x)$ which
are purely ``outgoing'' as $|x|\to \infty,$ i.e.,  the functions that are moving away from
the potential barrier in  both directions. For a perturbation of a given spin weight,
$s,$ the quasinormal modes are labeled by the multipole number, $l,$ and the overtone 
number $n.$

Typically, the form of the potential function practically excludes the possibility of finding 
an exact solution of Eq.(\ref{schred}) and consequently one is forced to adopt the numerical or
approximate methods. One of the most popular approximate (semianalytic) methods is the
Iyer-Schutz-Will approach~\cite{wkb0,wkb1} and its generalization to the sixth-order 
WKB~\cite{Konoplya6}. The Iyer-Schutz-Will modification of the WKB method in its original form
allows for potentials with closely lying classical turning points. 
It can be achieved by constructing the approximate solution in the region between the turning
points, which is expressible in terms of the parabolic cylinder function, and matching them with
the approximate WKB solutions in the exterior regions. 
For example, in Ref.~\cite{wkb1} the third order WKB solutions are asymptotically matched with
the solutions constructed for the 6th-order Taylor approximation of the potential. It is
interesting that the results presented in~\cite{wkb1} have been reconstructed by Zaslavskii, who
used textbook formulas describing the energy levels of the anharmonic oscillator and by Galt'sov
and Matukhin. The Iyer-Schutz-Will method has been subsequently extended to higher order in 
Ref.~\cite{Konoplya6}. 

A natural question that arises in this context is whether one could construct a better
approximation yielding more accurate results. In Ref.~\cite{ja} (henceforth Paper I) 
it was shown that the answer to this question is affirmative. The strategy adopted in
that paper is to extend the order of the WKB terms  (and consequently the order of
the Taylor series approximation of the potential) and construct the diagonal Pad\'e
approximants of the formal series for $\omega^{2}.$ It has been shown that within
the domain of validity of the method, the diagonal Pad\'e transforms 
$\mathcal{P}_{6}^{6}$ and $\mathcal{P}_{7}^{6}$ are always in excellent agreement 
with the numerical (exact) results constructed for the perturbed Schwarzschild 
and Reissner-Nordstr\"om black holes. Moreover, it has been shown that the method 
works well even for the fundamental low-lying modes of the perturbed Tangherlini black
hole. Unfortunately the analytic calculations of the higher order terms of the WKB
approximation are very complicated and each next order requires more and more time.
Recently, in a calculation lasting tens of hours, we have calculated all the terms up to
16th order, which allow  to extend our method to $\mathcal{P}_{8}^{8}$ transforms.
However, it would be interesting to extend and verify our method in the regime of 
higher order WKB. There is little hope that  this can be done analytically for a general
potential as the number of terms as well as their complexity grows fast with the order.
On the other hand, however, one expects that there should be massive simplifications 
when the calculations are performed for a given potential with prescribed multipole 
and overtone numbers. Another way out of the impasse would be the extension of Zaslavskii's
approach to really high orders. Recently, in the interesting and important 
paper devoted to Borel summation of the analytically continued Pad\'e transforms, 
Hatsuda~\cite{hatsuda} propounded to make use of the efficient Mathematica package
written by Sulejmanpasic and Ünsal~\cite{sulejman} (see also Ref.~\cite{BW} and the
references therein). This package allows calculation of the energy corrections 
of the one-dimensional system with an arbitrary locally-harmonic potential. This, 
in turn, leads us to the method  advocated by Zaslavskii.

In this paper we report on a further extension of our method by using a combination 
of different procedures, which will reveal some  unexpected features of the WKB
approximation and will shed fresh light on the problem of overtones. The main emphasis 
is put on accuracy that can be achieved using relatively simple techniques and, as in 
Paper I, we are looking for a simple, WKB-based approximation that can be used, 
practically without any changes, for a wide class of metrics. This black-box
nature of the method, with the effective potential taken as the input and the accurate 
frequencies of the the quasinormal modes obtained as the output, should be considered as
its great advantage.  The results presented in this paper raise the question of the
limitations of the method. The usual interpretation is that the method works well for
low overtones and  the approximation gets progressively better with increasing $l.$ 
However, the results presented in this paper show that one can achieve great accuracy
even for higher overtones at the expense of increasing of the order of the perturbation
series.

\section{The method} 
Although the derivation of the equations of the standard Iyer-Schutz-Will method 
is rather complicated it leads, surprisingly, to quite simple final result. Since 
the method and its generalizations have been presented in details in 
Refs.~\cite{wkb0,wkb1,ja} we shall omit most of its technical aspects here. 

The formula relating the complex frequencies of the quasinormal modes and 
the derivatives 
of $Q(x)$ at $x=x_{0}$ can be written in a compact form $(n=0,1,2,...)$
\begin{equation}
 \frac{ i Q_{0}}{\sqrt{  2Q''_{0}}\varepsilon} -\sum_{k=2}^{N} \varepsilon^{k-1}\Lambda_{k}= n+ \frac{1}{2},
 \label{master}
\end{equation} 
where
$x_{0}$ is a point at which the potential has its maximum,
each $\Lambda_{k}$ is combination of the derivatives of $Q(x)$ calculated at $x_{0}$ and 
$\varepsilon$ is the expansion parameter that helps to keep track of the order of terms. 
The complexity of $\Lambda_{k}$ terms increases with $k$ and the number of the constituents 
for $k \leq 16$ is given in Table I.

\begingroup\squeezetable
\begin{table}
 \caption{\label{tabb1} The number of terms in $\Lambda_{k}$ for $ 2 \leq k \leq 16. $}
\begin{ruledtabular}
\begin{tabular}{cccccccccccccccc} 
$\Lambda_{n}$ & $\Lambda_{2}$ & $\Lambda_{3}$ & $\Lambda_{4}$ & $\Lambda_{5}$ & $\Lambda_{6}$ & 
$\Lambda_{7}$ & $\Lambda_{8}$ & $\Lambda_{9}$ & $\Lambda_{10}$ & $\Lambda_{11}$ & $\Lambda_{12}$ & $\Lambda_{13}$ & 
$\Lambda_{14}$ & $\Lambda_{15}$ & $\Lambda_{16}$\\  \hline
Number of terms & 6 & 20 &  55 & 132 & 294 & 616 & 1215 & 2310 & 4235 & 7524 & 13026& 22050 & 36540 & 59488 & 95268\\ 
\end{tabular}
 \end{ruledtabular}
 \end{table}
 \endgroup
 
Now, let us assume that the potential does not depend on $\omega.$  In a standard approach 
one puts $\varepsilon =1$ and  for a given potential  solves Eq.~\ref{master} with respect to 
$\omega.$ That means that the $\Lambda_{k}$ terms are summed. On the other hand, however, 
we do not know in advance if taking into account additional $\Lambda$ terms will improve the
quality of the results. Moreover, one cannot exclude the possibility that the improvement is 
just accidental.  To remedy this problem it has been proposed to treat the expression for the 
quasinormal frequencies as the formal perturbation expansion
\begin{equation}
 \omega^{2} = V(x_{0})-i \left(n+\frac{1}{2} \right)  \sqrt{  2Q''_{0}} \varepsilon -i 
 \sqrt{  2Q''_{0}} \sum_{i=2}^{N} \varepsilon^{j} \Lambda_{j}
 \equiv V(x_{0})  + \sum_{i=1}^{N} \varepsilon^{i} \tilde{\Lambda}_{i}
 \label{omm}
\end{equation}
and instead of summing the terms in the right hand side of~(\ref{omm}) (which is a bad strategy) to construct 
the Pad\'e approximants~\cite{ja}. As is well known the Pad\'e approximants associated
with a formal power series $\sum a_{k} x^{k}$ are defined as the unique rational functions $\mathcal{P}_{N}^{M}(x)$ 
of degree $N$ in the denominator and $M$ in the numerator 
satisfying 
\begin{equation}
 \mathcal{P}_{N}^{M}(x) - \sum_{k=0}^{M+N} a_{k}x^{k} ={ \cal{O}}(x^{M+N+1})
\end{equation}

In  our previous paper we have extended the WKB-based calculations to 
the 13th-order, i.e., we have calculated all $\Lambda_{k}$ up to $k=13$ and for any 
given multipole  $l$ and the overtone number $n$ we used the Pad\'e summation~\cite{CarlB}. 
It has been demonstrated that for the so-far best documented quasinormal frequencies of the
Schwarzschild and Reissner-Nordstr\"om black holes the diagonal Pad\'e transforms $\mathcal{P}_{6}^{6}$
and $\mathcal{P}^{6}_{7}$ are always in very good agreement with the exact numerical results. 
Of course the approximation has its own limitations mainly related to the limitations of the WKB theory.
On the other hand, the deviations of the complex frequencies  obtained within
the framework of the improved method  from the numerical results are smaller than the analogous
results obtained with the aid of competing (approximate) approaches. In this paper we extend
the analyses of Ref.~\cite{ja} in two ways: first we will construct the $\Lambda_{k}$ functions
for $k\leq 16$ and subsequently use the Wynn epsilon algorithm for convergence acceleration~\cite{wynn,brezinski}. 
Details of the algorithm are relegated to Appendix. Although
the formulas describing  $\Lambda_{k}$ are rather complicated the calculations can be
substantially accelerated by the choice of the suitable strategy. As the construction
of the functions $\Lambda_{k}$ has been described in detail in Refs.~\cite{wkb0,wkb1,ja} 
we will not repeat it here. 
Unfortunately, if one is interested in really high orders of the WKB approximation 
the analytic approach briefly described above is hard, if not impossible, to use. Although 
its numerical variant would certainly simplify and speed up the calculations here we will
follow a different path. We will make use of the Zaslavskii approach in which the resonant
scattering problem is converted into the bound state problem of the anharmonic oscillator. 
The simplest method to do so is to exploit a formal equivalence between the master equation~(\ref{schred}) 
with the function $Q(x)$ expanded in a Taylor series about the point
$x_{0}$ at which it has a maximum and the equation that describes quantum-mechanical 
anharmonic oscillator. Indeed, reducing the latter problem to the simple application 
of the standard perturbation theory, one has
\begin{equation}
 \left( H_{0} +\tilde{V} \right)\psi = E \psi,
\end{equation}
where 
\begin{equation}
 H_{0} = -\frac{1}{2} \frac{d^{2}}{dx^{2}} + \frac{1}{4} Q_{0}^{(2)} x^{2},
\end{equation}
\begin{equation}
 \tilde{V} = \frac{1}{2}\sum_{k=3}^{N}  \varepsilon^{\frac{k}{2}-1}Q_{0}^{(k)} \frac{x^{k}}{k!} 
\end{equation}
and $Q_{0}^{(k)}$ is $k$-th derivative of $Q(x)$ at $x_{0}.$ Now, calculating the perturbative 
corrections to the energy levels, multiplying the thus obtained results by the factor 
$\sqrt{2/Q^{(2)}_{0}}$ and finally making the substitution $\varepsilon \to  i \varepsilon$ one obtains precisely 
the $\Lambda_{j}$ functions of Eq.~(\ref{omm}). 

As have been observed by Zaslavskii, the first two nontrivial 
correction terms, i.e. the Iyer-Will result, can easily be calculated using almost entirely textbook results~\cite{LL}. 
Unfortunately, the textbook-based method is rather inefficient in the higher-order calculations. Recently, in a very 
interesting paper Sulejmanpasic and and Ünsal~\cite{sulejman}
extended the method developed by Bender and Wu~\cite{BW} (and originally applied to the simple  anharmonic oscillator) 
to the arbitrary locally harmonic potential. The paper is accompanied with
the very efficient computer algebra package, allowing construction of the really high orders 
of the perturbation expansion. Although the practical applicability of the package for the analytic  calculations 
of the general functions $\Lambda_{k}$ is limited to first few orders
its real power lies in the numerics\footnote{We have checked that for $k\leq 5$ the $\Lambda_{k}$  calculated with the aid of the 
BenderWu package are precisely the same as the functions $\Lambda_{k}$  constructed within the framework 
of the Iyer-Will approach. The equality of the results for $6 \leq k\leq 16$  has been verified numerically. }
Our strategy is as follows: First  we calculate the complex 
frequencies of the black hole normal modes using the general $\Lambda_{k}$ $(k \leq 16)$ constructed within 
the framework of the Iyer-Will approach
and subsequently, for a given harmonic and overtone numbers, we calculate numerically the high orders of 
parameters $\Lambda_{k}.$  (Note the slightly abused notation). The choice of the maximal $k$ used 
in this paper depends on the type of  the  modes, 
but it never exceeds $k=700.$ Having computed all the necessary $\Lambda_{k}$ we construct the Pad\'e approximants 
of Eq.~(\ref{omm}) treated as a formal polynomial of $\varepsilon.$ We show that one can obtain amazingly 
accurate results even in the situations typically considered as too hard for the WKB-based methods. 

We conclude this section with a few observations on the approach proposed in 
Ref.~\cite{hatsuda}. First it should be observed that although the Pad\'e transforms,
$\tilde{\mathcal{P}}_{N}^{N}$ of the series $\sum_{k} a_{k} x^{k}/k!$ can easily be constructed, 
the calculation of the integral transform of $\tilde{\mathcal{P}}_{N}^{N},$ denoted here by  $\mathcal{B}_{N}^{N}$
\begin{equation}
 \mathcal{B}_{N}^{N} = \int_{0}^{\infty} e^{-\xi} \tilde{\mathcal{P}}_{N}^{N}(\xi x) d\xi,
 \label{lapl}
\end{equation}
could be time consuming, especially for $k >100.$ Moreover, the calculations of 
$\mathcal{B}_{N}^{N}$ for $k > 100$ 
suffer (at least in our implementation of the algorithm) from numerical instabilities, and, consequently,  
one has to be very careful in choosing the calculational strategy and in interpreting the thus obtained results.

\section{Quasinormal modes}
\subsection{The Schwarzschild black hole}

In this section we shall construct the quasinormal modes of the perturbed Schwarzschild 
black hole. Although the Pad\'e summation is our method of choice, whenever in doubt, we
will also use the Borel-Pad\'e summation~\cite{hatsuda}. Moreover, all the cases presented 
in this section have been calculated using the continued fraction method.  It should be 
stressed that the Pad\'e transforms work extremely well and our results are in perfect 
agreement with the  continued fraction method. We have also corrected a few erroneous results, 
which have appeared in the literature.

\begin{figure}
\centering
\includegraphics[width=12cm]{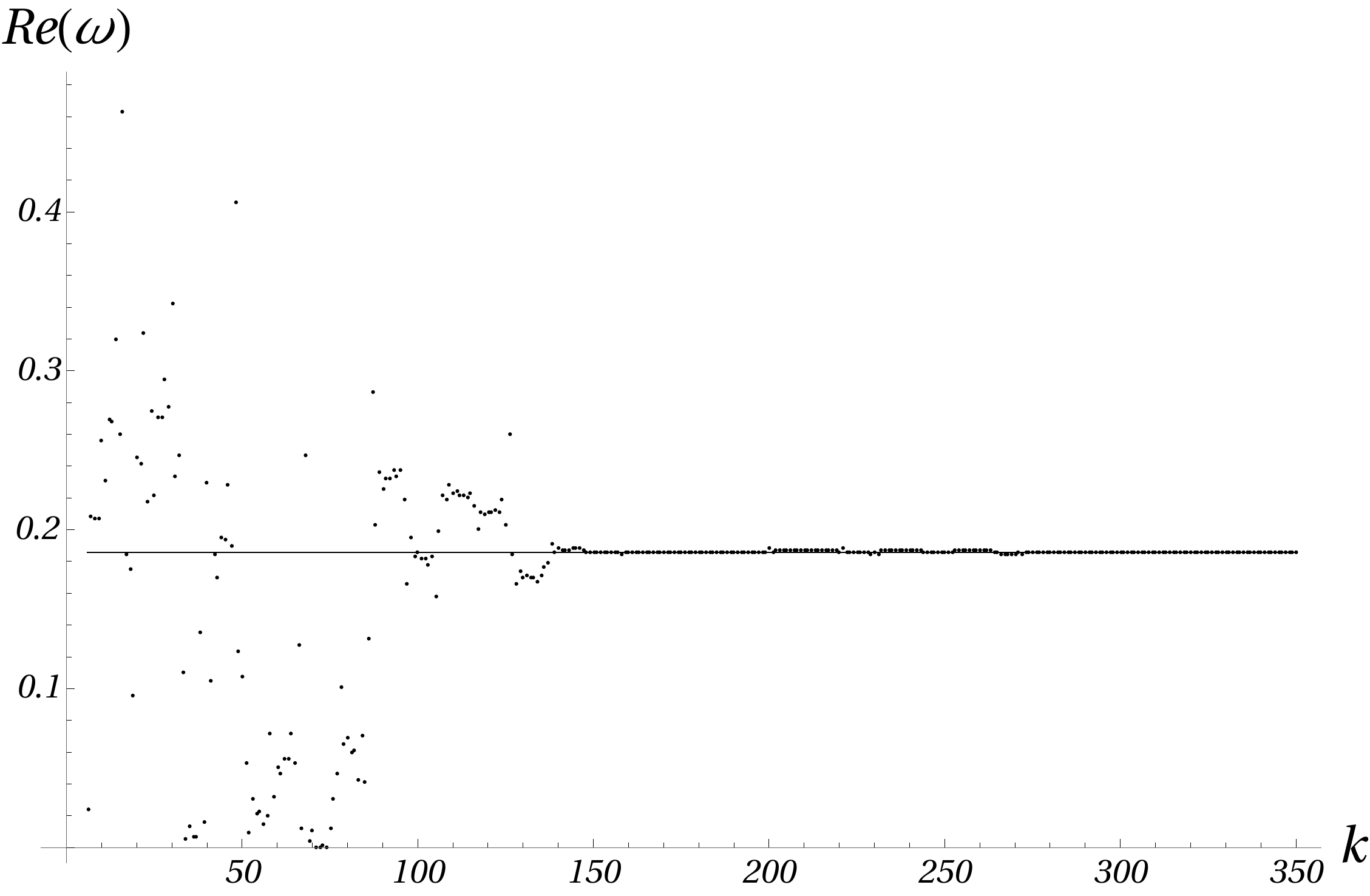}
\caption{The real part of the quasinormal frequency of the odd gravitational perturbation with  $l=2, $ $n=7$ 
calculated for 345  Pad\'e transforms $\mathcal{P}_{k}^{k}, $  $6 \leq k \leq 350.$ The horizontal thin solid 
line represents the exact numerical result. }
\label{fig:7re}
\end{figure}

\begin{figure}
\centering
\includegraphics[width=12cm]{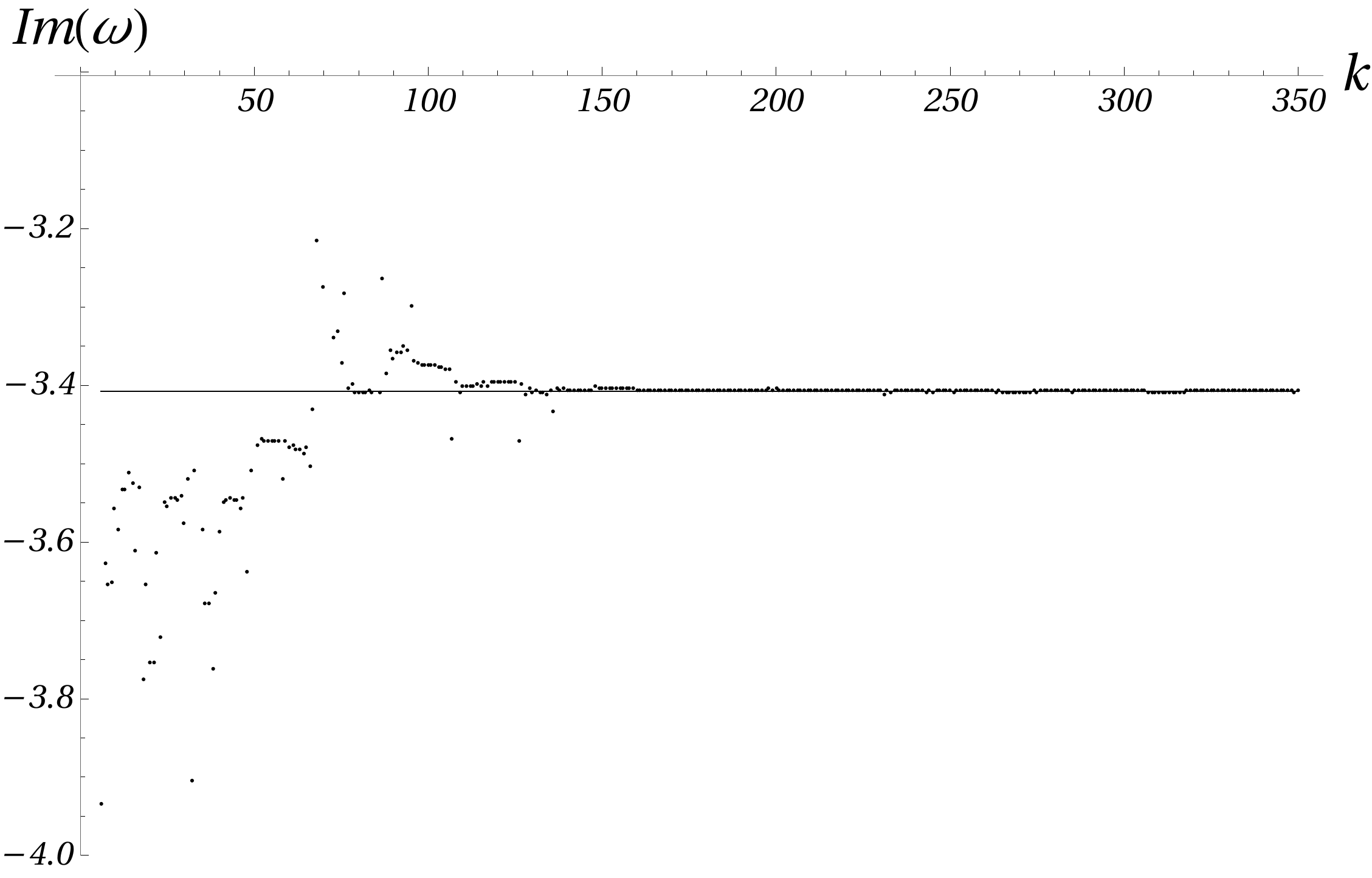}
\caption{The imaginary part of the quasinormal frequency of the odd gravitational 
perturbation with  $l=2, $
$n=7$ calculated for 345  Pad\'e transforms $\mathcal{P}_{k}^{k}, $  $6 \leq k \leq 350.$ 
The horizontal thin solid line represents the exact numerical result.}
\label{fig:7im}
\end{figure}

First let us consider the odd-parity (Regge-Wheeler) potential
\begin{equation}
 V(r_{\ast}) = \left(1 -\frac{1}{r} \right)\left(\frac{L}{r^{2}} + \frac{1-s^2}{r^{3}} \right),
\end{equation}
where $L = l(l+1)$ and $s =0, 1$ or $2$ for the scalar, vector and gravitational perturbations, 
respectively. The problem of construction of the complex frequencies 
of the quasinormal modes reduces to a simple application of the formula~(\ref{omm}).
The results of our calculations are tabulated in Tables~\ref{tabA0}-\ref{ta5} and the accurate 
numerical results presented at the top of each column are taken from 
Ref.~\cite{nils}. Generally speaking all the results follow the same pattern which is 
clearly visible in Figs.~\ref{fig:7re} and \ref{fig:7im}. Each point represent the real 
(Fig.\ref{fig:7re}) or imaginary (Fig.\ref{fig:7im}) part of the complex frequency $\omega$ 
constructed for a given ${\mathcal{P}_{k}^{k}}$ for a $s=2,$ $l=2$ and $n=7$. Inspection 
of the Figures shows that even  for $k=100$ the spread of $\omega$ is quite big, and that the result 
seems to stabilize starting with $k =150.$ The horizontal line represents the exact numerical result. 
The analogous behavior of lower overtones is similar to that discussed above with the one reservation: 
the spread of $\omega$ is smaller and stabilization starts for smaller values of $k.$ The lesson 
that follows from this test case is that one should be cautious with accepting the results based 
on only a few first $\Lambda$ terms. Similarly, the equality of a few consecutive results does 
not necessarily mean that the final stable result is approached.
The above discussion raises two questions: Does this pattern hold for higher overtones, and 
is it possible to reconstruct the algebraically special modes this way. Although a complete
answer to this questions is beyond our understanding, we can, nevertheless, make some
observations. First, it seems impossible to construct the algebraically special modes, say 
the gravitational mode (2,8), by any of the WKB-based methods. For an interesting discussion
see Andersson's paper~\cite{nils}. The rest of this section is devoted to the first question. 

The common wisdom regarding the applicability of the WKB method to
the quasinormal mode problem is that it gives, at best, only approximate results, 
satisfactorily reproduces only the fundamental modes and works really great for $l \gg 1.$
Moreover, the frequencies of the overtones can be calculated with some confidence  only for 
$n \lesssim l$ and rapidly deteriorates  with the index number.
And finally, summing the first few terms  of the perturbative series for $\omega^{2}$ can 
give more accurate results.  Below we shall show that some of the above limitations may 
be weakened or even overcome except the last point which is generally untrue.

To analyze the quality of the approximation in more details it
is convenient to define deviation of the real part of the frequency
\begin{equation}
 \Delta^{(r)}(\omega_{k}) = \frac{\Re(\omega_{k})-\Re(\omega_{num})}{\Re(\omega_{num})} 100\%
\end{equation}
and the deviation of its imaginary part
\begin{equation}
 \Delta^{(i)}(\omega_{k}) = \frac{\Im(\omega_{k})-\Im(\omega_{num})}{\Im(\omega_{num})} 100 \%,
\end{equation}
where $\omega_{k}$ is the approximate complex frequency of the quasinormal mode and 
$\omega_{num}$ is its accurate numerical value. Now, consider for example the scalar (odd) 
(0,10) mode. Our calculations show that the deviations of $\omega$ constructed from 
${\mathcal{P}_{350}^{350}}$ are  $\Delta^{(r)} (\omega) = 6.37 \times 10^{-2}\%$ and 
$\Delta^{(i)}(\omega)  = 6.14 \times 10^{-4}\%, $ respectively. Although the accuracy 
of this result is quite good, and we expect that for the higher-order  transforms it would
be even better, the calculation becomes progressively more difficult, and, consequently,
there is a natural limit of applicability of the method\footnote{The total time of calculations 
rapidly grows with $k.$}. On the other hand, for a given order of 
the Pad\'e sum the accuracy of the approximation increases with the harmonic index $l$ 
and deteriorates with $n.$

To simplify our discussion we shall denote  $l,$ $n$ mode by $(l,n)$  and the Borel-Pad\'e 
sum constructed from the diagonal Pad\'e transform $\tilde{\mathcal{P} }_{N}^{N}$ by 
$\mathcal{B}_{N}^{N}.$ Since the numerical calculations reported in Ref.~\cite{nils}
were limited to an accuracy of nine decimal places (with the exception of the 
scalar (0,0), (0,1), (1,0) and (1,1) modes) our results, for comparison, have been
rounded-off accordingly. (In the tables the results of Ref.~\cite{nils} have been
multiplied by 2, whereas for numerical comparison our results have been divided by 2).

First, let us consider the lowest scalar mode $(0,0).$ This mode is somewhat mysterious 
as there is a discrepancy between the exact numerical results and the results obtained within 
the framework of the Pad\'e summation (see Tab.~\ref{tabA0}).
The frequency calculated by Andersson is $ 0.2209086 - 0.2097886 i,$
whereas the Pad\'e summation gives the result that differs at the 6th and 5th decimal
places  for the real and imaginary part of $\omega,$ respectively. Moreover, to the order
considered in this paper, the result seems to stabilize with the rather high accuracy of 
16 decimal places. Additionally we have calculated the complex frequency of $(0,0)$ mode 
using the Borel-Pad\'e summation and for 
$\mathcal{B}_{249}^{249}$ and  $\mathcal{B}_{250}^{250}$ the results obtained are in concord 
with the Pad\'e transforms rather than the result presented in Ref.~\cite{nils}. Indeed, 
the results obtained within the framework of the both methods agree to 12 decimal places. 
It is possible that the WKB-based methods give the convergent result that differs form 
the exact numerical one.
On the other hand however, it can be seen from Tab.~\ref{tabbA1} that for 
the mode $(0,1)$ one has $\Delta^{(r)} = \Delta^{(i)} =0$ and the complex 
frequencies of the  higher overtones are
amazingly accurate (Tabs.~\ref{tabbA2}-\ref{tabbA5}). To solve this
problem we have calculated $\omega$ using the method of continued fraction 
with the Wynn algorithm to accelerate convergence of the approximants.
Our code (a very unsophisticated one) gives for the fundamental frequency
precisely the same result as the Pad\'e summation to the accuracy quoted, i.e., the 
results are identical to 16 decimal places. This equality, although expected, 
is really impressive, especially taking into account the deep differences in the 
computational strategies.  For a better confidence we have computed the frequency 
of this mode without the series acceleration in the continued fraction method and
obtained precisely the same result. There is also a small discrepancy in the real part 
of the scalar mode (1,0). Inspection of Tab.~\ref{tabA0} shows that we have precisely the
same situation as the one discussed above. We believe that our results are correct.

It should be emphasized that although the results are tabulated only for a few exemplary
values of $k$ we have calculated $\omega$ for all $k$ satisfying $6 \leq k \leq k_{max},$
where $k_{max}$ is the maximal order of the Pad\'e transform involved in the
calculations. That, in turn, means that determination of the Pad\'e transform 
$\mathcal{P}_{k}^{k}$ requires knowledge of the perturbation series to the order $2k.$
Now, let us consider the scalar modes $l=1$ listed in the right columns of 
Tabs.~\ref{tabA0}-\ref{tabbA5}. Our calculations show that the Pad\'e approximants 
yield frequencies that are in a perfect agreement ($\Delta^{(r)} =\Delta^{(i)}=0$)  
with the numerical calculations reported in Ref.~\cite{nils}.

The Pad\'e approximants for the electromagnetic modes are also accurate. Indeed, 
the deviations of the vector $l=1$ modes (see Tables~\ref{ta0}-\ref{ta5}) are 
identically zero for all considered cases. It should be noted however, that the result
for the scalar (1,0) mode (see Tab.~\ref{ta0}) presented in Ref.~\cite{nils} differs 
from our calculations and to establish which result is the correct one we have compared
the frequency with the result obtained using the continued fraction method. Inspection 
of the table shows that the Pad\'e transforms and the continued fraction method give
precisely the same $\omega,$ and, consequently, we believe that our result is correct.

Finally, for the gravitational modes listed in the right columns of 
Tabs.~\ref{ta0}-\ref{ta5}, one finds $|\Delta^{(r)} |\propto 10^{-7}\,\%$ and 
$|\Delta^{(r)} | \propto 10^{-6 }\,\%$ for $n=4$ and $n=5,$  respectively.
Similarly, for the deviations of the imaginary parts one has  
$|\Delta^{(i)} |= \propto 10^{-6}\,\%$  for  $n=4$ and $n=5.$  The remaining 
results exactly reproduce the numerically calculated frequencies.

We have focused on the Regge-Wheeler potential. One expects that the quality of the 
approximation  grows with $l.$ Similar calculations have been carried out for the Zerilli potential, 
and, as expected, the obtained results are even better. Unfortunately, the calculations for the 
even parity modes, because of the form of the Zerilli potential, are more complicated. As they
add little to the discussion they will not be presented here.

\begin{center}
\begin{table}
 \caption{\label{tabA0} The complex frequencies of the scalar quasinormal modes for $s=0,$ $l=0,$ $n=0$ 
 (left column) and  $s=0,$ $l=1,$ $n=0$ (right column). The Pad\'e 
 approximants, $\mathcal{P}_{k}^{k},$ are calculated for $k =6,8,50,100,150,200, 250$ and $300.$ 
 The last row is calculated within the framework of the continued fraction method. }
\begin{tabular}{rcc}
 $k$ & $l = 0,$ $n =0$ &  $l = 1,$ $n =0$\\ \hline
 & $        0.2209086 - 0.2097886 i$  &  $0.5858722 - 0.19532 i$ \\ \hline
   6 & $0.2226575720860749-0.2097705214916321 i$  & $0.5858723081529635-0.1953200537758777 i$  \\
   8 & $0.2209467125923617-0.2099078086678199 i$  & $0.5858722675722460-0.1953199825012170 i$  \\
  50 & $0.2209098826757620-0.2097914422244537 i$  & $0.5858722665345654-0.1953199778271564 i$  \\
 100 & $0.2209098781648214-0.2097914341697581 i$  & $0.5858722665345654-0.1953199778271564 i$  \\
 150 & $0.2209098781608154-0.2097914341737470 i$  & $0.5858722665345654-0.1953199778271564 i$  \\
 200 & $0.2209098781608393-0.2097914341737620 i$  & $0.5858722665345654-0.1953199778271564 i$  \\
 250 & $0.2209098781608394-0.2097914341737619 i$  & $0.5858722665345654-0.1953199778271564 i$  \\ 
 300 & $0.2209098781608394-0.2097914341737619 i$  & $0.5858722665345654-0.1953199778271564 i$  \\
 CF  & $0.2209098781608394-0.2097914341737619 i$  & $0.5858722665345654-0.1953199778271564 i$  \\
\end{tabular}
\end{table}
\end{center}

\begin{center}
\begin{table}
 \caption{\label{tabbA1} The complex frequencies of the scalar quasinormal modes. The Pad\'e 
 approximants, $\mathcal{P}_{k}^{k},$ are calculated for $k =6,8,50,100,150,200$ and $250.$}
\begin{tabular}{rcc}
 $k$ & $l = 0,$ $n =1$ &  $l = 1,$ $n =1$\\ \hline
 &$0.1722338 - 0.6961048 i$ & $0.5288973 - 0.61251478 i$ \\ \hline
   6 & $0.1739745319353183-0.6959893881331765 i$   &$0.5288861492458066-0.6125171735439730 i$  \\  
   8 & $0.1757654368519873-0.6957757352543876 i$   &$ 0.5288974070506602-0.6125174288621156 i$ \\
  50 & $0.1722542901637571-0.6961018341205277 i$   &$ 0.5288973012096589-0.6125147831180967 i$ \\
 100 & $0.1722345014978213-0.6961051951489169 i$   &$ 0.5288973012096651-0.6125147831180942 i$ \\
 150 & $0.1722338624300778-0.6961048943217377 i$   &$ 0.5288973012096651-0.6125147831180942 i$ \\
 200 & $0.1722338376996682-0.6961048946120284 i$   &$ 0.5288973012096651-0.6125147831180942 i$ \\
 250 & $0.1722338367597048-0.6961048936507108 i$   &$ 0.5288973012096651-0.6125147831180942 i$ \\
\end{tabular}
\end{table}
\end{center}

\begin{center}
\begin{table}
 \caption{\label{tabbA2} The complex frequencies of the scalar quasinormal modes. The Pad\'e
 approximants, $\mathcal{P}_{k}^{k},$ are calculated for $k =6,8,50,100,150,200$ and $250.$}
\begin{tabular}{rcc}
 $k$ & $l = 0,$ $n = 2$ &  $l = 1,$ $n = 2$\\ \hline
 &$0.151483872 - 1.20215718 i$  & $45907867 - 1.08026685 i$  \\ \hline
   6 & $0.1501610934536142-1.2183646740525384 i$   & $0.4584542419793649-1.0801478608208415 i$ \\   
   8 & $0.1638975530635953-1.2177361392404707 i$   & $0.4588978123676309-1.0803319265755279 i$ \\
  50 & $0.1525086624622595-1.2030996120120083 i$   & $0.4590786697505832-1.0802668500082639 i$ \\
 100 & $0.1514835672117383-1.2021637621538090 i$   & $0.4590786698626058-1.0802668500382169 i$ \\
 150 & $0.1514855070708692-1.2021579660190725 i$   & $0.4590786698626033-1.0802668500382144 i$ \\
 200 & $0.1514840139830334-1.2021573800732860 i$   & $0.4590786698626033-1.0802668500382144 i$ \\
 250 & $0.1514838833327049-1.2021571932998896 i$   & $0.4590786698626033-1.0802668500382144 i$ \\
\end{tabular}
\end{table}
\end{center}

\begin{center}
\begin{table}
 \caption{\label{tabbA3} The complex frequencies of the scalar quasinormal modes. The Pad\'e
 approximants, $\mathcal{P}_{k}^{k},$ are calculated for $k =6,8,50,100,150,200$ and $250.$}
\begin{tabular}{rcc}
\hline
 $k$ & $l = 0,$ $n = 3$ &  $l = 1,$ $n = 3$\\ \hline
 & $0.140820276 - 1.707354636 i$ & $0.406516772 - 1.576595646 i$ \\ \hline
  6 & $0.138060417273466-1.732103754671567 i$     & $0.405851086039425-1.577770124846531 i$ \\
  8 & $0.137670535339324-1.731506857378750 i$     & $0.406197938547983-1.576954341222488 i$ \\
 50 & $0.142414546944632-1.706681174348947 i$     & $0.406516784404676-1.576595635301799 i$ \\
100 & $0.140982067753947-1.707290791968112 i$     & $0.406516772362319-1.576595645568015 i$ \\
150 & $0.140847814828334-1.707387507328528 i$     & $0.406516772366920-1.576595645562394 i$ \\
200 & $0.140821949939888-1.707357238989158 i$     & $0.406516772366927-1.576595645562396 i$ \\
250 & $0.140820464201770-1.707355063984697 i$     & $0.406516772366927-1.576595645562396 i$ \\ \hline
\end{tabular}
\end{table}
\end{center}

\begin{center}
\begin{table}
\caption{\label{tabbA4} The complex frequencies of the scalar quasinormal modes. The Pad\'e
 approximants, $\mathcal{P}_{k}^{k},$ are calculated for $k =6,8,50,100,150,200$ and $250.$}
\begin{tabular}{rcc}
\hline
 $k$ & $l = 0,$ $n = 4$ &  $l = 1,$ $n = 4$\\ \hline
& $0.134148608 - 2.21126376 i$ &  $0.37021804 - 2.081524226 i$ \\ \hline
  6 & $0.124765034276338-2.229022763903854 i$  &   $0.369834774192733-2.084205198945555 i$ \\ 
  8 & $0.131144317225359-2.239048678679388 i$  &   $0.369767198720716-2.081642311885022 i$ \\
 50 & $0.140769758034688-2.209616863307420 i$  &   $0.370217997741403-2.081524178098750 i$ \\
100 & $0.133932901572174-2.212104617277194 i$  &   $0.370218041051879-2.081524225583624 i$ \\
150 & $0.134187347043185-2.211290539046596 i$  &   $0.370218040690568-2.081524225634893 i$ \\
200 & $0.134168230131617-2.211267909887815 i$  &   $0.370218040690412-2.081524225635177 i$ \\
250 & $0.134149076601310-2.211261806257428 i$  &   $0.370218040690404-2.081524225635138 i$ \\
   \hline
\end{tabular}
\end{table}
\end{center}

\begin{center}
\begin{table}
\caption{\label{tabbA5} The complex frequencies of the scalar quasinormal modes. The Pad\'e
 approximants, $\mathcal{P}_{k}^{k},$ are calculated for $k =6,8,50,100,150,200,250,300$ and $350.$}
\begin{tabular}{rcc}
\hline
 $k$ & $l = 0,$ $n = 5$ &  $l = 1,$ $n = 5$\\ \hline
& $ 0.129483444 - 2.71427893 i$ & $ 0.344153622 - 2.588239396 i$ \\ \hline
   6 & $0.122524417275065-2.726977710003108 i$  & $0.349813936639104-2.592216963795157 i$ \\
   8 & $0.127292811291479-2.746960760117273 i$  & $0.340849558720576-2.589050268997814 i$ \\
  50 & $0.146613305792775-2.691720611629606 i$  & $0.344154713036727-2.588239612402011 i$ \\
 100 & $0.133076394282346-2.714297398099659 i$  & $0.344153626253300-2.588239388487938 i$ \\
 150 & $0.129736760525904-2.714220815430054 i$  & $0.344153622489611-2.588239396618380 i$ \\
 200 & $0.129545864551587-2.714271452225676 i$  & $0.344153622500862-2.588239396622716 i$ \\
 250 & $0.129501756504425-2.714283142375501 i$  & $0.344153622501171-2.588239396622020 i$ \\
 300 & $0.129484950919583-2.714282779384554 i$  & $0.344153622501291-2.588239396621862 i$ \\
 350 & $0.129484811363228-2.714277877939901 i$  & $0.344153622501297-2.588239396621861 i$ \\   
	\hline
\end{tabular}
\end{table}
\end{center}

\begin{center}
\begin{table}
\caption{\label{ta0} The complex frequencies of the electromagnetic (left column) and the 
gravitational (right column) quasinormal modes. The Pad\'e approximants,   
$\mathcal{P}_{k}^{k},$ are calculated for $k =6,8,50,100,150,200,$ and $250.$ The last row is 
calculated within the framework of the continued fractions method.}
\begin{tabular}{rcc}
\hline
 $k$ & $s=1$, $l = 1,$ $n = 0$ & $s=2,$  $l = 2,$ $n = 0$\\ \hline
     & $0.496526544 - 0.184975418 i$ & $0.747343368 - 0.17792463 i$ \\ \hline
   6 & $0.4965264765276178-0.1849753180356798 i$  & $0.7473429777005388-0.1779414742208077 i$\\
   8 & $0.4965265309901696-0.1849754375379761 i$  & $0.7473517597949078-0.1779529416751695 i$ \\
  50 & $0.4965265283562174-0.1849754359058844 i$  & $0.7473433688364812-0.1779246313780083 i$ \\
 100 & $0.4965265283562174-0.1849754359058844 i$  & $0.7473433688360837-0.1779246313778714 i$ \\
 150 & $0.4965265283562174-0.1849754359058844 i$  & $0.7473433688360837-0.1779246313778714 i$ \\
 200 & $0.4965265283562174-0.1849754359058844 i$  & $0.7473433688360837-0.1779246313778714 i$ \\
 250 & $0.4965265283562174-0.1849754359058844 i$  & $0.7473433688360837-0.1779246313778714 i$ \\
 CF  & $0.4965265283562174-0.1849754359058844 i$  & $0.7473433688360837-0.1779246313778714 i$ \\
	\hline
\end{tabular}
\end{table}
\end{center}

\begin{center}
\begin{table}
\caption{\label{ta1} The complex frequencies of the electromagnetic (left column) and the 
gravitational (right column) quasinormal modes. The Pad\'e
 approximants, $\mathcal{P}_{k}^{k},$ are calculated for $k =6,8,50,100,150,200,$ and $250$}
\begin{tabular}{rcc}
\hline
 $k$ & $s=1$, $l = 1,$ $n = 1$ & $s=2,$  $l = 2,$ $n = 1$\\ \hline
     &  $0.42903084 - 0.587335292 i$ & $0.693421994 - 0.54782975 i$ \\ \hline
    6 & $0.4290303098689009-0.5873349677300929 i$   & $0.6925703481161420-0.5478549212072782 i$   \\     
    8 & $0.4290305267101144-0.5873351193512594 i$   & $0.6931350943803045-0.5477990964454646 i$   \\   
   50 & $0.4290308391272459-0.5873352910914822 i$   & $0.6934219986356213-0.5478297513079539 i$   \\  
  100 & $0.4290308391272117-0.5873352910914573 i$   & $0.6934219937602342-0.5478297505799500 i$   \\   
  150 & $0.4290308391272117-0.5873352910914573 i$   & $0.6934219937583319-0.5478297505824732 i$   \\   
  200 & $0.4290308391272117-0.5873352910914573 i$   & $0.6934219937583269-0.5478297505824696 i$   \\   
  250 & $0.4290308391272117-0.5873352910914573 i$   & $0.6934219937583269-0.5478297505824696 i$   \\   
	\hline
\end{tabular}
\end{table}
\end{center}

\begin{center}
\begin{table}
\caption{\label{ta2} The complex frequencies of the electromagnetic (left column) and the 
gravitational (right column) quasinormal modes. The Pad\'e
 approximants, $\mathcal{P}_{k}^{k},$ are calculated for $k =6,8,50,100,150,200,$ and $250$}
\begin{tabular}{rcc}
\hline
 $k$ & $s=1$, $l = 1,$ $n = 2$ & $s=2,$  $l = 2,$ $n = 2$\\ \hline
      & $0.349547136 - 1.050375198 i$ & $0.60210691 - 0.956553966 i$ \\ \hline
    6 & $0.3489671476223886-1.0496242138899032 i$    & $0.5983320642239965-0.9571047257361685 i$ \\   
    8 & $0.3495311626539206-1.0501579826696264 i$    & $0.6010669702662552-0.9565205248055362 i$ \\   
   50 & $0.3495471345091076-1.0503751985469316 i$    & $0.6021058209312484-0.9565564091949364 i$ \\   
  100 & $0.3495471352140788-1.0503751987176345 i$    & $0.6021069091277328-0.9565539641411267 i$ \\   
  150 & $0.3495471352140217-1.0503751987176476 i$    & $0.6021069092310929-0.9565539664324743 i$ \\   
  200 & $0.3495471352140216-1.0503751987176475 i$    & $0.6021069092254648-0.9565539664461968 i$ \\   
  250 & $0.3495471352140216-1.0503751987176475 i$    & $0.6021069092247347-0.9565539664461398 i$ \\   
	\hline
\end{tabular}
\end{table}
\end{center}

\begin{center}
\begin{table}
\caption{\label{ta3} The complex frequencies of the electromagnetic (left column) and the 
gravitational (right column) quasinormal modes. The Pad\'e
 approximants, $\mathcal{P}_{k}^{k},$ are calculated for $k =6,8,50,100,150,200,$ and $250$}
\begin{tabular}{rcc}
\hline
 $k$ & $s=1$, $l = 1,$ $n = 3$ & $s=2,$  $l = 2,$ $n = 3$\\ \hline
      & $0.292353398 - 1.543817848 i$    & $0.503009924 - 1.410296404 i$ \\ \hline
   6  & $0.288963705877476-1.544282449993818 i$   & $0.497642737535928-1.414461228437683 i$ \\ 
   8  & $0.291218230899165-1.542799891892294 i$   & $0.502401150125607-1.403417125878443 i$ \\
  50  & $0.292353425313164-1.543817856808729 i$   & $0.503101540951305-1.410129166751418 i$ \\
 100  & $0.292353398843248-1.543817848040974 i$   & $0.503010351175324-1.410295016204194 i$ \\
 150  & $0.292353398833784-1.543817847996815 i$   & $0.503009929112050-1.410296418310432 i$ \\
 200  & $0.292353398833987-1.543817847996173 i$   & $0.503009925121021-1.410296404044260 i$ \\
 250  & $0.292353398834001-1.543817847996167 i$   & $0.503009924395821-1.410296404890974 i$ \\
        \hline
\end{tabular}
\end{table}
\end{center}

\begin{center}
\begin{table}
\caption{\label{ta4} The complex frequencies of the electromagnetic (left column) and the 
gravitational (right column) quasinormal modes. The Pad\'e
 approximants, $\mathcal{P}_{k}^{k},$ are calculated for $k =6,8,50,100,150,200,$ and $250$}
\begin{tabular}{rcc}
\hline
 $k$ & $s=1$, $l = 1,$ $n = 4$ & $s=2,$  $l = 2,$ $n = 4$\\ \hline
     & $0.253108292 - 2.045100568 i$ & 0.41502916 - 1.893689782 i \\ \hline
   6 & $0.248727904828560-2.051639959255668 i$  & $0.419396702975745-1.887928109921840 i$ \\
   8 & $0.249369705013264-2.043584186827301 i$	& $0.362047466802911-1.868264500927380 i$ \\
  50 & $0.253108599503254-2.045101202339157 i$	& $0.415360773226012-1.890523875905394 i$ \\
 100 & $0.253108285283085-2.045100567418578 i$	& $0.415038826730706-1.893625733917996 i$ \\
 150 & $0.253108292027439-2.045100567596560 i$	& $0.415026686988849-1.893690158005340 i$ \\
 200 & $0.253108291983075-2.045100567606178 i$	& $0.415029155910513-1.893689969992049 i$ \\
 250 & $0.253108291980756-2.045100567606751 i$	& $0.415029158009993-1.893689749075853 i$ \\
	\hline
\end{tabular}
\end{table}
\end{center}

\begin{center}
\begin{table}
\caption{\label{ta5} The complex frequencies of the electromagnetic (left column) and the 
gravitational (right column) quasinormal modes. The Pad\'e
 approximants, $\mathcal{P}_{k}^{k},$ are calculated for $k =6,8,50,100,150,200, 250, 300$ and $350$}
\begin{tabular}{rcc}
\hline
 $k$ & $s=1$, $l = 1,$ $n = 5$ & $s=2,$  $l = 2,$ $n = 5$\\ \hline
      & $0.224505582 - 2.547851238 i$& $0.338598806 - 2.391216108 i$ \\ \hline
    6 & $0.222128242094723-2.559732389855251 i$   & $0.306732391439588-2.297611394962681 i$ \\     
    8 & $0.217300021185279-2.546769260540304 i$   & $0.244507176638539-2.455516877615405 i$ \\     
   50 & $0.224508451763544-2.547853939163384 i$   & $0.310122138984459-2.392572467682007 i$ \\     
  100 & $0.224505537892275-2.547851205268914 i$   & $0.337381549712206-2.391221582182717 i$ \\     
  150 & $0.224505582024767-2.547851238125507 i$   & $0.338677506810238-2.391243385562942 i$ \\     
  200 & $0.224505581675340-2.547851238746054 i$   & $0.338592848394283-2.391225800860825 i$ \\     
  250 & $0.224505581611109-2.547851238755523 i$   & $0.338596687693972-2.391217123833891 i$ \\ 
  300 & $0.224505581599715-2.547851238757362 i$   & $0.338598570386376-2.391216091649921 i$ \\
  350 & $0.224505581599435-2.547851238756736 i$   & $0.338598780167286-2.391216068179005 i$ \\
	\hline 
\end{tabular}
\end{table}
\end{center}

\subsection{Reissner-Nordstr\"om black hole}

We conclude this section with a brief discussion of the perturbations of the Reissner-Nordstr\"om 
black hole. The normal modes satisfy the differential equation (\ref{schred}) with the potential of the form
\begin{equation}
V(x) = \left(1- \frac{1}{r}+ \frac{Q^{2}}{r^{2}}\right) 
\left( \frac{l(l+1)}{x^{2}} -\frac{q_{2s-3}}{x^{3}}+ \frac{4 Q^{2}}{x^{4}}  \right),
\end{equation}
where the parameters $q_{\pm}$ are defined as
\begin{equation}
q_{\pm 1} = \frac{1}{2} \left( 3 \pm \sqrt{9 + 16 q^{2} (l-1)(l+2)} \right),
\end{equation}
and $s =1$ and $s=2$ for the electromagnetic and gravitational perturbations, respectively. 
It should be emphasized that all the results  cited in Ref.~\cite{hatsuda} have been 
easily reproduced by constructing the appropriate Pad\'e transforms of the perturbation series.
Since the results of our calculations follow the common pattern we shall discuss only one
exemplary configuration with $Q=1/5$  and $l=2$ (see Tab.~\ref{rn1}). First, we have found 
that in order to calculate the quasinormal frequency of the (2,0) mode with the accuracy 
quoted in~\cite{hatsuda}, we need $\mathcal{P}_{k}^{k}$ for $k \geq 158.$ Moreover,
starting with $k =206$ the result stabilizes with an accuracy of (at least) 24 decimal places.
It should be noted that the Borel summation applied to $\tilde{\mathcal{P}}_{250}^{250}$  gives 
precisely the same result. For the higher overtones both methods give very close results.

We conclude this subsection with a comment that also relates to our previous discussion.
First, it should be noted that calculation of the Pad\'e transforms or making use 
of the Wynn algorithm takes only a small fraction of the total time. Most of the
computation time is spent on construction of the $\Lambda_{k}$ parameters. Moreover,
in order to avoid zero divisors one has to retain as many digits as necessary, that
may considerably slow down the calculation. Nevertheless, they can be completed in a reasonable 
time. On the other hand, calculation of the integral transforms~(\ref{lapl})
may be tricky and time consuming.

\begin{center}
\begin{table}
\caption{\label{rn1} The complex frequencies of the gravitational quasinormal $l=2$ modes of the 
Reissner-Nordstr\"om black hole. $Q=1/5.$ }
\begin{tabular}{rcc}
\hline
 mode & method & $\omega$  \\ \hline
$n=0$ & $\mathcal{P}_{250}^{250}$ & $0.756873775755127489560243 - 0.178796227993424581484091 i$ \\
$n=0$ & $\mathcal{B}_{250}^{250}$ & $0.756873775755127489560243 - 0.178796227993424581484091 i$ \\
$n=1$ & $\mathcal{P}_{250}^{250}$ & $0.703455061389587189545552 - 0.550248607591426397610300 i$ \\
$n=1$ & $\mathcal{P}_{300}^{300}$ & $0.703455061389587189590323 - 0.550248607591426397588809 i$ \\
$n=1$ & $\mathcal{B}_{250}^{250}$ & $0.703455061389587189595369 - 0.550248607591426397589084 i$ \\
$n=2$ & $\mathcal{P}_{250}^{250}$ & $0.612848616773723964811455 - 0.959880899685908114593243 i$ \\
$n=2$ & $\mathcal{P}_{350}^{350}$ & $0.612848616773725485129278 - 0.959880899685896111782658 i$ \\                                          
$n=2$ & $\mathcal{B}_{350}^{350}$ & $0.612848616773725500883098 - 0.959880899685896118184149 i$ \\
\hline 
\end{tabular}
\end{table}
\end{center}

\section{Final remarks}
\label{fin}

We have developed a simple strategy for calculating quasinormal modes of the black holes by
extending the approach propounded in Paper I. The basic ingredients of the method are
the generalization of the Iyer-Will technique to higher-order terms of the WKB approximation
and subsequent summation of the thus obtained perturbation series with the aid of the Pad\'e
transforms. Although the general form of the functions $\Lambda_{k}$ are known only for
$k\leq 16,$ the higher-order $\Lambda_{k}$ parameters (for a given harmonic and overtone number)
can easily be calculated numerically. Once the $\Lambda_{k}$ are known, the complex quasinormal
frequency can be obtained from the Pad\'e transform of the perturbation series. Our method has
been successfully used to calculate the quasinormal modes of the Schwarzschild and Reissner-
Nordstr\"om black holes. Indeed, we have demonstrated that the results are amazingly accurate,
even for the overtones. For example, the fundamental gravitational $l=2$ mode calculated using 
the continued fractions method agrees with the result of our approach to 24 decimal places, 
which is really impressive. We are unable to answer the question if the present technique works well
for the highly-damped modes. The answer would require calculations of the Pad\'e 
transforms far beyond the limit assumed in this paper.  However, all the cases that have been
considered here seem to obey a simple rule: the more terms of the perturbation series are retained
and transformed the better the final result is obtained, although the improvement may be 
slow\footnote{It should be noted that the convergence of the continued fraction method is also
slow.}.

We have demonstrated that our method is accurate and useful and we believe that it is also the
simplest one. It can be applied to a wide class of problems practically without any change. 
Indeed, its simple black-box structure, with the (expanded) effective potential as the input and 
the accurate quasinormal frequencies as the output is certainly its attractive feature.
Moreover, it is relatively fast and all the steps  are numerically stable. 
We believe that it also would be the first method of choice in many applications 
and experimentation, as  summing up the terms of the perturbation series is generally a bad
strategy. The Mathematica notebooks with examples and explanations can be obtained from the 
first author upon request.

\appendix*
\section{}
\begin{figure}
\centering
\includegraphics[width=10cm]{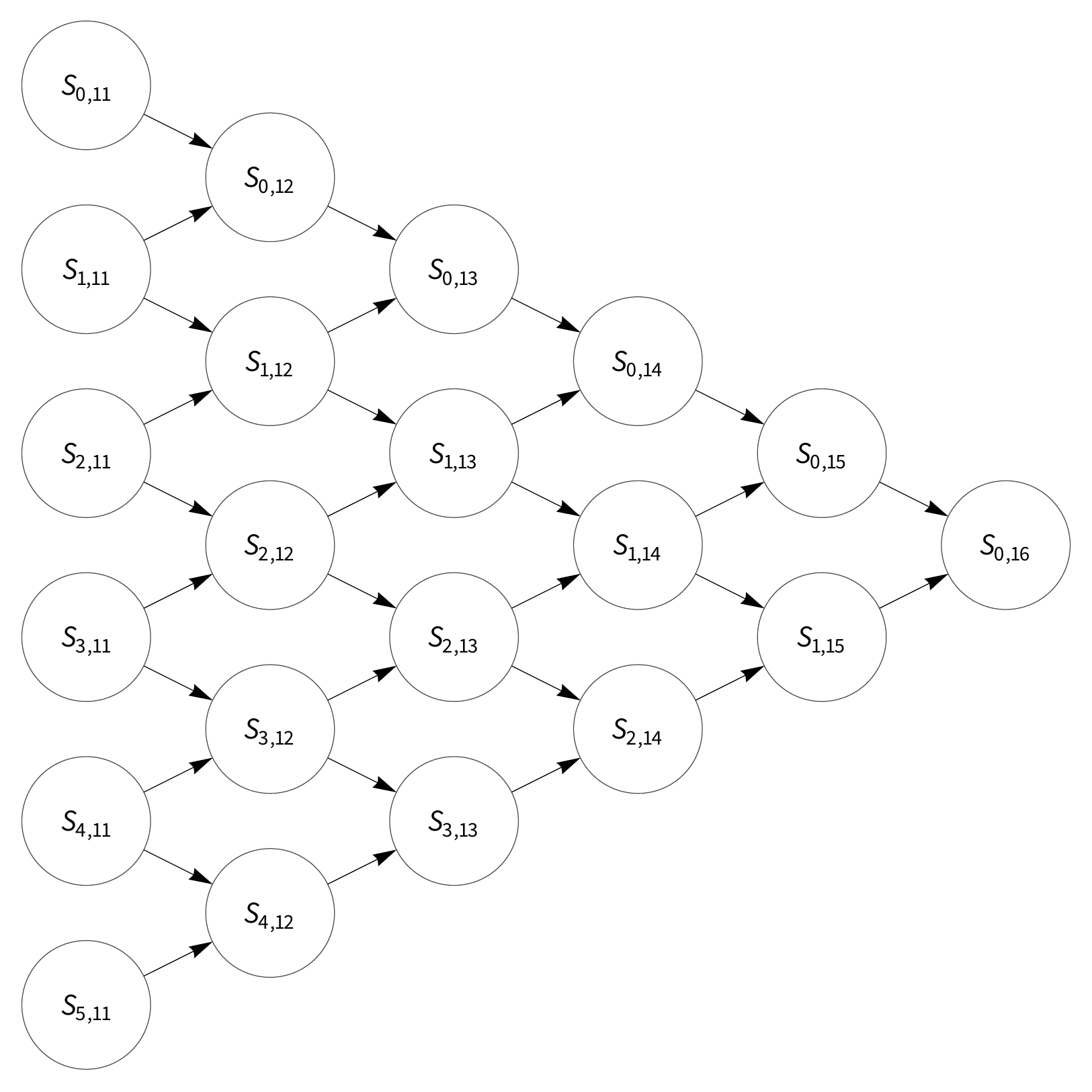}
\caption{The few last columns of the $\epsilon$ transforms of the 16th order problem.
It should be noted that although only the last entry inherits all previous $S$-terms,
each $S_{j,k}$ does inherit all $\tilde{\Lambda}_{i}$  up to $j+k.$}
\label{fig:diamond}
\end{figure}

In this Appendix, we describe briefly the Wynn epsilon algorithm heavily used in this paper.
First, let us introduce our notation. Let $S =(s_{0},s_{1},..., s_{i} \in \mathbb{C})$ 
be a sequence of the partial sums (in our case it is the sequence of the WKB approximants). 
The Wynn algorithm consists of the initialization phase in which we put
\begin{equation}
  S_{j,0} = s_{j}
\end{equation}
and define
\begin{equation}
 S_{j,-1} =0,
\end{equation}
and the iterative phase 
\begin{equation}
 S_{j,k+1} = S_{j+1,k-1} + \frac{1}{S_{j+1,k} -S_{j,k}}.
\end{equation}
In the Wynn $\epsilon$ algorithm we are interested in terms with even $k.$ Typically, 
the terms with odd $k$ grow and the approximants are labeled by even $k.$  A few last columns 
of $S_{j,k},$  where $S_{j,0}$ ($j=0,1,...,n_{max}$) are the partial sums of the terms in (\ref{omm})
are schematically shown in Fig.~\ref{fig:diamond}.
There is a deep connection between the $S_{j,k} $ terms and the  Pad\'e approximants.
Indeed, the entries $S_{j,2k}$ in the columns labeled by the even index are equal to 
the Pad\'e approximants
\begin{equation}
S_{j,2 k} =P_{k}^{j+k} ,
\end{equation}  
provided the usual assumptions of the existence of the approximants $P$  and 
nonexistence of the zero divisors in the construction of the $S$ transforms are satisfied.
It should be noted that the remaining approximants of the Pad\'e table should be
constructed independently. Specifically, of the diagonal approximants $P_{k}^{k}$ and 
$P_{k+1}^{k}$ only the former can be calculated this way. 

The Wynn algorithm is simple and very efficient, however, since the elements of 
the odd columns grow there is a real danger of numerical instabilities. This can be
cured by retaining the sufficient number of digits in the calculations. The diamond-like
structure of the Wynn algorithm makes it very easy to use.


\end{document}